\documentclass[a4paper,11pt]{article}
\usepackage{jinstpub} 
\usepackage{lineno}

\proceeding{LIDINE 2023: LIght Detection In Noble Elements\\
 Sep 20-22, 2023\\ Madrid, Spain}

\title{Light detection and Cosmic Rejection in the ICARUS LArTPC at Fermilab}

\collaboration[c]{on behalf of the ICARUS collaboration}

\author[1]{A. Heggestuen}
\affiliation[1]{Colorado State University,\\
Fort Collins, CO 80523, USA}

\emailAdd{aheggest@colostate.edu}

\abstract{The ICARUS-T600 detector is a 760-ton Liquid Argon Time Projection Chamber (LArTPC) currently operating at Fermilab as the Far Detector in the Short Baseline Neutrino (SBN) program. The SBN program is composed of three LArTPCs with a central goal of testing the sterile neutrino hypothesis. After operating for 3-years in the Gran Sasso Underground Laboratory, the ICARUS detector was shipped to CERN where it was outfitted with 360 8” Photomultiplier Tubes (PMTs) for a new optical detection system. The PMT system detects fast scintillation light from charged particles interacting in the Liquid Argon, generating the trigger signal for the full detector and allows 3D reconstruction of events. Now operating at shallow depth, the detector is exposed to a high flux of cosmic rays that can fake neutrino interactions. To mitigate this effect a Cosmic Ray Tagger (CRT) and a 3-meter-thick concrete were installed. Precise timing information from both the PMT and CRT subsystems can help to identify whether an interaction originated from inside or outside of the ICARUS cryostat. This paper reviews a method for cosmogenic background reduction and timing calibration of the CRT and PMT light detection systems in ICARUS.}

\begin{document}
\maketitle
\flushbottom

\label{sec:intro}
\section{Introduction}
The Short Baseline Neutrino (SBN) Program \cite{sbnproposal} is geared to help resolve several experimental anomalies observed in neutrino physics that hint towards the presence of an additional neutrino state. Comprised of three Liquid Argon Time Projection Chambers (LArTPCs) placed at various distances along the $\sim$800 MeV Booster Neutrino Beam (BNB) beamline at Fermilab, this program will provide the most sensitive searches to date for sterile neutrinos at the eV-mass scale. The Short Baseline Near Detector (SBND), located just 110 m from the beam target, is expected to record over a million neutrino interactions per year and is currently being commissioned. MicroBooNE was situated 470 m from the target and collected data from 2015-2021, recording hundreds of thousands of neutrinos from the same BNB beam. ICARUS is located 600 m from the target and is the first large scale operating LArTPC, containing 760 tons of ultra-pure Liquid Argon. 

LArTPC technology was first proposed by Carlo Rubbia in 1977 \cite{Rubbia:1977zz} and developed by the ICARUS collaboration for over 25 years. When a particle interacts with the LAr, both scintillation light and ionization charge is produced. The scintillation light is detected by an array of photo-multiplier tubes (PMTs). A uniform electric field is applied drifting the ionization electrons to three parallel wire planes placed 3 mm apart and located in front of the PMTs. The wire planes give three 2D views of ionizing particles interacting in the LAr, while the PMT gives nanosecond level timing of the interaction, allowing for fine-grained 3D reconstruction of particle trajectories with mm-level spatial resolution in a large detector volume.
\label{sec:intro}
\label{sec:icarus}
\section{The ICARUS-T600 detector}
ICARUS first operated in the Gran Sasso underground laboratory situated under 1,400 m of rock studying neutrinos from CERN. After a successful 3-year run, the detector was shipped to CERN for upgrades where it was outfitted with new TPC electronics and a new light detection system. 

In 2017, ICARUS moved to Fermilab where its now operating at shallow depth as the far detector in the SBN program. The high flux of cosmogenic particles at Earth's surface poses new challenges for this highly sensitive detector as these background particles can interact in the argon and mimic the primary neutrino signal. On average, $\sim$ 13 cosmic muon tracks are expected to cross the detector volumn in the 1 ms TPC drift readout. To help mitigate the cosmic ray flux, a 3-m thick concrete overburden was installed above the detector to stop particles before they reach the detector (see \cite{Behera:2021iap} for a detailed study on the impact of the overburden). In addition, a Cosmic Ray Tagging (CRT) system completely surrounds the detector medium ($\sim$~1,000~$m^2$ coverage) with two layers of fiber embedded plastic scintillator to tag the remaining incoming charged particles. 

The full ICARUS detector is composed of two symmetric TPCs, each approximately 3.6 m high, 3.9 m wide and 19.9 m long. 
Each TPC has a central cathode with a maximum drift distance of 1.5 m (drift time $\sim$0.96 ms at the nominal 500 V/cm electric field). The two anode planes in each TPC are made of three parallel wire planes (horizontal, +60$^{\circ}$ and -60$^{\circ}$), amounting to $\sim$~54,000 total wires. 
360 8” Hamamatsu R5912-MOD PMTs are coated with tetra-phenyl butadiene (TPB) to convert the VUV photons to visible light. The PMT system uses laser calibration system for gain equalization, timing and monitoring. 
In addition to the BNB beam, the ICARUS detector is situated at a unique position where it is also exposed to Neutrinos at the Main Injector (NuMI) beam (6$^{\circ}$~off-axis), a higher energy neutrino beam provided by Fermilab. ICARUS finished commissioning in Spring 2022, allowing for the first physics data run exposed to the BNB and NuMI beams in June 2022. 
\label{sec:icarus}
\label{sec:pmt}
\section{Light detection system and trigger}
The ICARUS light detection system detects scintillation light from charged particles interacting in the LAr. The fast scintillation light detected by the PMTs yields reconstruction of the position of the interaction along the charge drift direction and is used for the trigger of the full detector. 
While the charge drift time in LAr is about a millisecond, the PMTs have a few nanosecond resolution providing a precise timestamp of when the interaction happened. 
The PMT signals are generated as low-voltage differential signaling (LVDS) logic outputs are first digitized and discriminated in pairs. If five pairs of PMTs have signal above a set threshold (Majority logic) within a 6-meter longitudinal slice (30~PMTs~x~2~sides~of~cathode) in coincidence with one of the beam spills (BNB:~1.6~$\mu$s, NuMI:~9.6~$\mu$s), a global trigger signal is produced. This global trigger signal then gets transmitted to the rest of the detector (TPC and CRT subsystems, remaining PMTs) indicating the presence of the neutrino beam at the detector. All data triggered by this signal are recorded (on-beam trigger). 

The current trigger rate in ICARUS at Fermilab $\sim$~0.7~Hz (0.3~Hz~BNB, 0.15~Hz~NuMI, 0.25~Hz~off-beam). The trigger efficiency is under investigation on data: above 200 MeV desposited energy there is full detection efficiency, but a lower efficiency for low energy cosmic rays is observed. This motivates a complimentary system based on \textit{Adder} boards. These \textit{Adders} take an analog sum of 15 PMTs and can help identify events with plenty of light but not enough fired PMTs. 

\begin{figure}
    \centering
    \includegraphics[width=1\textwidth]{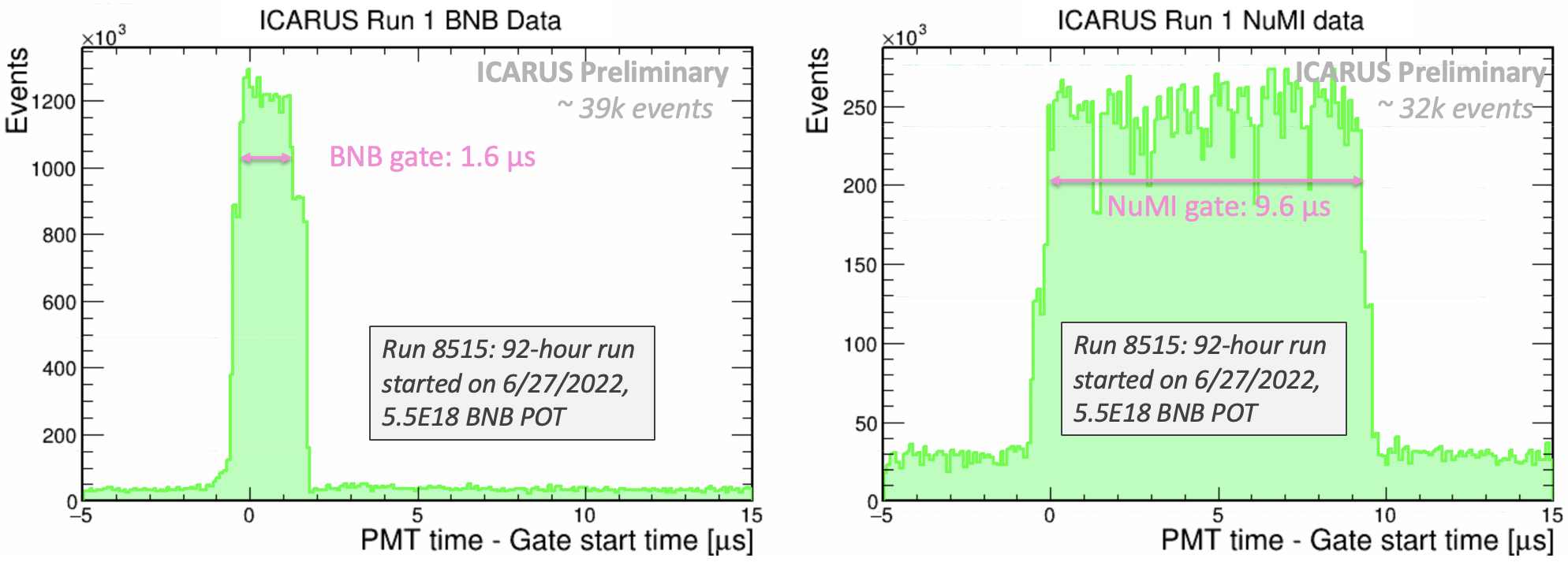}
    \caption{PMT scintillation timing signal with respect to the opening of the neutrino beam gates for ICARUS beam data shows excess PMT light over the standard cosmic background rate. Left: BNB Data. Right: NuMI data.}
    \label{fig:PMTTime}
\end{figure}

Additional triggers exist for cosmogenic background studies for neutrino oscillation searches and calibration. For example, there is a trigger without any request on scintillation light (Minimum bias) and a trigger for outside the beam spill (off-beam), among others. Looking at the distribution of PMT scintillation signal timing with respect to the opening of the neutrino beam gates shows excess PMT light over the standard cosmic background rate, demonstrating the trigger performance (see \figurename{ \ref{fig:PMTTime}}).
\label{sec:pmt}

\section{Cosmic Ray Tagging system}
The ICARUS CRT System is composed of three subsystems: a top, side and bottom CRT. The top CRT is made up of new scintillator modules designed by CERN/INFN to intercept 80\% of cosmic muons entering the ICARUS LArTPCs. Each top CRT module has 16 scintillator strips (double layered: 8X and 8Y) with two fibers per strip readout at a single end with Silicon photomultipliers (SiPMs). The side CRT uses repurposed veto modules from the decommissioned MINOS experiment. Each module is 8-meters long and houses 20 scintillator stripes embedded with wavelength shifting fibers. Fibers are collected into snouts at either end of the modules and read out with SiPMs for a new optical readout system designed by CSU. The bottom CRT utilizes repurposed Double Chooz modules with a PMT based readout. 

The CAEN A1702 32-channel Silicon Photomultipliers Readout Front-End Board are used to provide readout and digitization of CRT Signals. These Front-End Boards (FEBs) provide amplification and shaping of the SiPM signals with a configurable discrimination threshold (0~-~50~photoelectrons). The FEBs can be daisy chained together and allow for triggering on an external signal or coincident signals in adjacent channels. 

\begin{figure}
    \centering
    \includegraphics[width=1\textwidth]{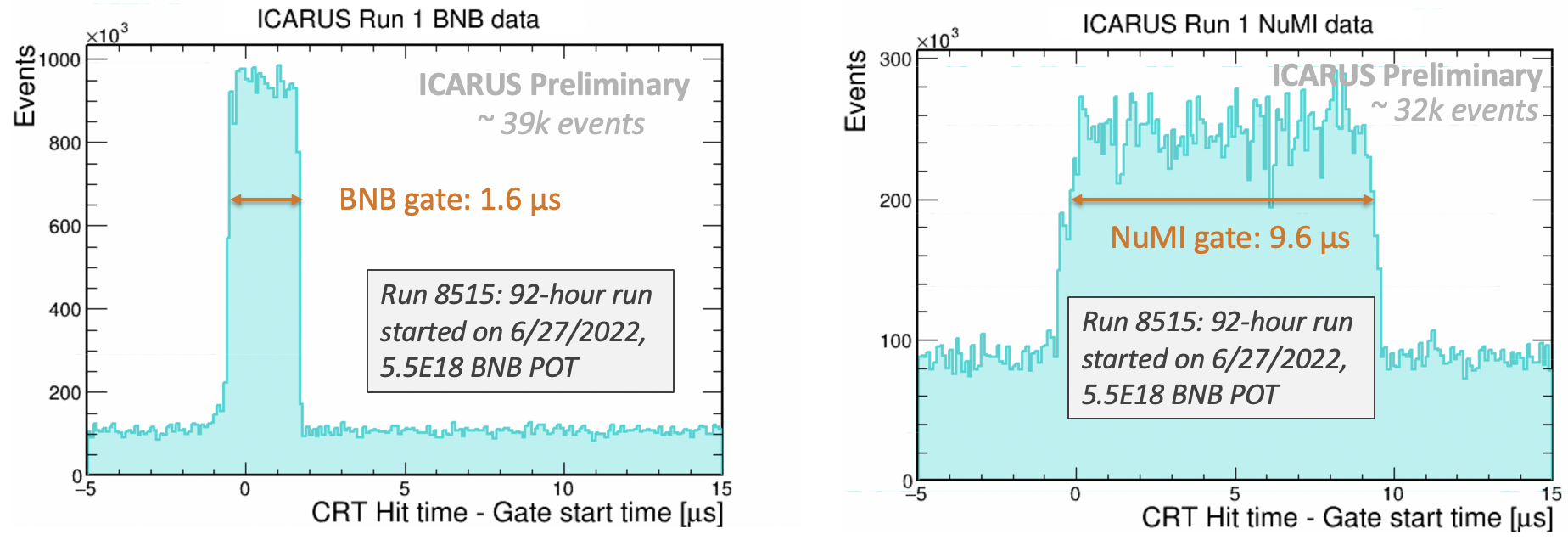}
    \caption{CRT Hit time with respect to the neutrino beam start time. Left: BNB Data. Right: NuMI data.}
    \label{fig:CRTTime}
\end{figure}

Each CRT FEB provides two internal counters, T0 and T1. The T0 counter receives input from a pulse per second (PPS) source, resetting every second to produce an absolute timestamp of every CRT interaction over threshold. The global trigger signal  indicating the presence of excess PMT light in coincidence with one of the neutrino beam spills, discussed in section \ref{sec:pmt}, is sent as an input to the T1 counter, providing a time reference to the beam timing inside the CRT system. 
Timing corrections are performed offline to adjust the absolute CRT Hit time for light propagation, cable delays and time walk with amplitude. Looking at the CRT timestamp with respect to the opening of the neutrino beam gates, excess hits associated with the BNB and NuMI beams are observed in figure \ref{fig:CRTTime}, similar to figure \ref{fig:PMTTime}. Each FEB has an internal resolution of $\sim$2.3~ns. 
\section{CRT - PMT matching for cosmic rejection}
Using ns-level timing information from the CRT Hits and PMT signals, muon tracks entering and exiting the ICARUS TPCs can be distinguished. This method 
looks for CRT Hits within 100 ns of a PMT flash and computes a simple time difference, $T_{CRT}-T_{PMT}$. If this value is negative, the CRT Hit was seen before the PMT activity (PMT flash) indicating that a particle entered the detector from outside. On the other hand, if $T_{CRT}-T_{PMT}$ is positive, the PMT light was detected before the CRT Hit and indicates that a particle first produced light inside the detector and then exited. 

\begin{figure}
    \centering
    \includegraphics[width=1\textwidth]{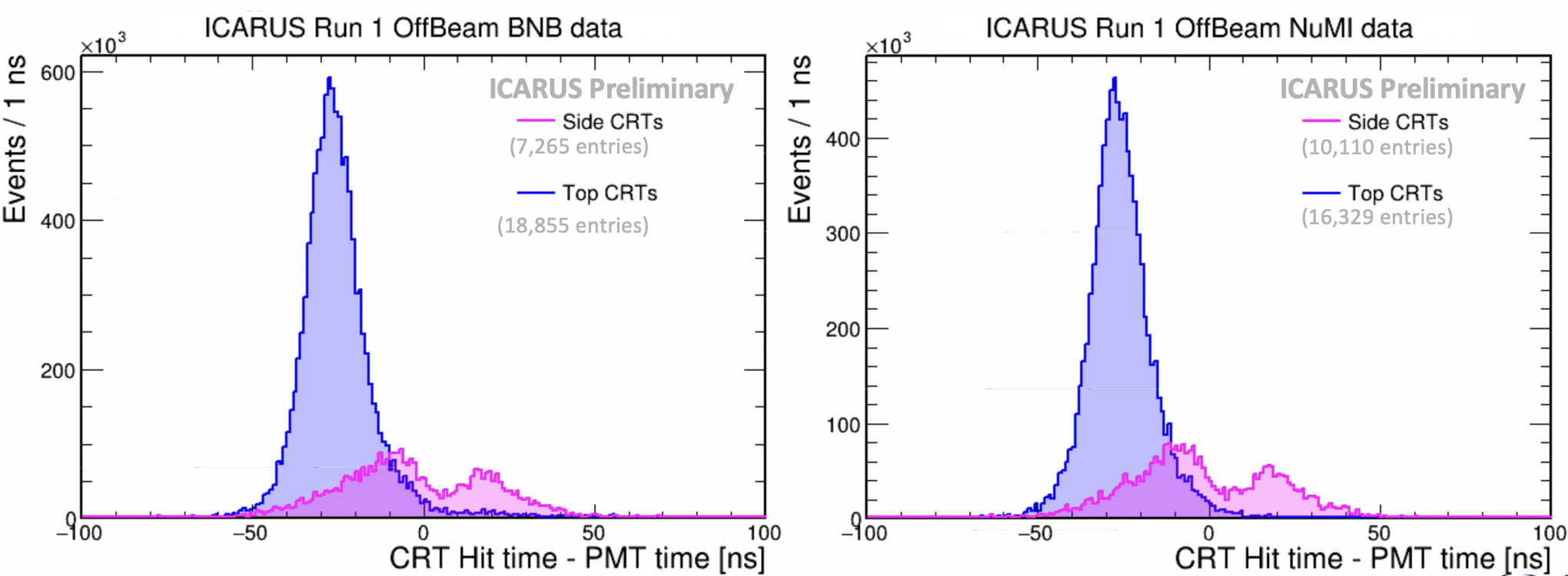}
    \caption{CRT Hit - PMT signal time difference for off-beam Run 1 data. PMT activity matched with a top CRT hit (blue) in comparison to PMT activity matched with a side CRT hit (pink). Left: BNB Data. Right: NuMI data.}
    \label{fig:CRTPMT}
\end{figure}

The CRT Hit - PMT flash time difference for off-beam BNB and NuMI ICARUS run 1 data is shown in figure \ref{fig:CRTPMT}. The distribution for top CRT matches is primary composed of cosmic tracks first hitting the top CRT and then producing a flash in the LAr, as expected due to the large top CRT coverage. The side CRT is able to tag cosmic tracks entering and exiting the detector volume, as shown with the two peaks in the distribution of side CRT matches in figure \ref{fig:CRTPMT}. 

\begin{figure}
    \centering
    \includegraphics[width=.5\textwidth]{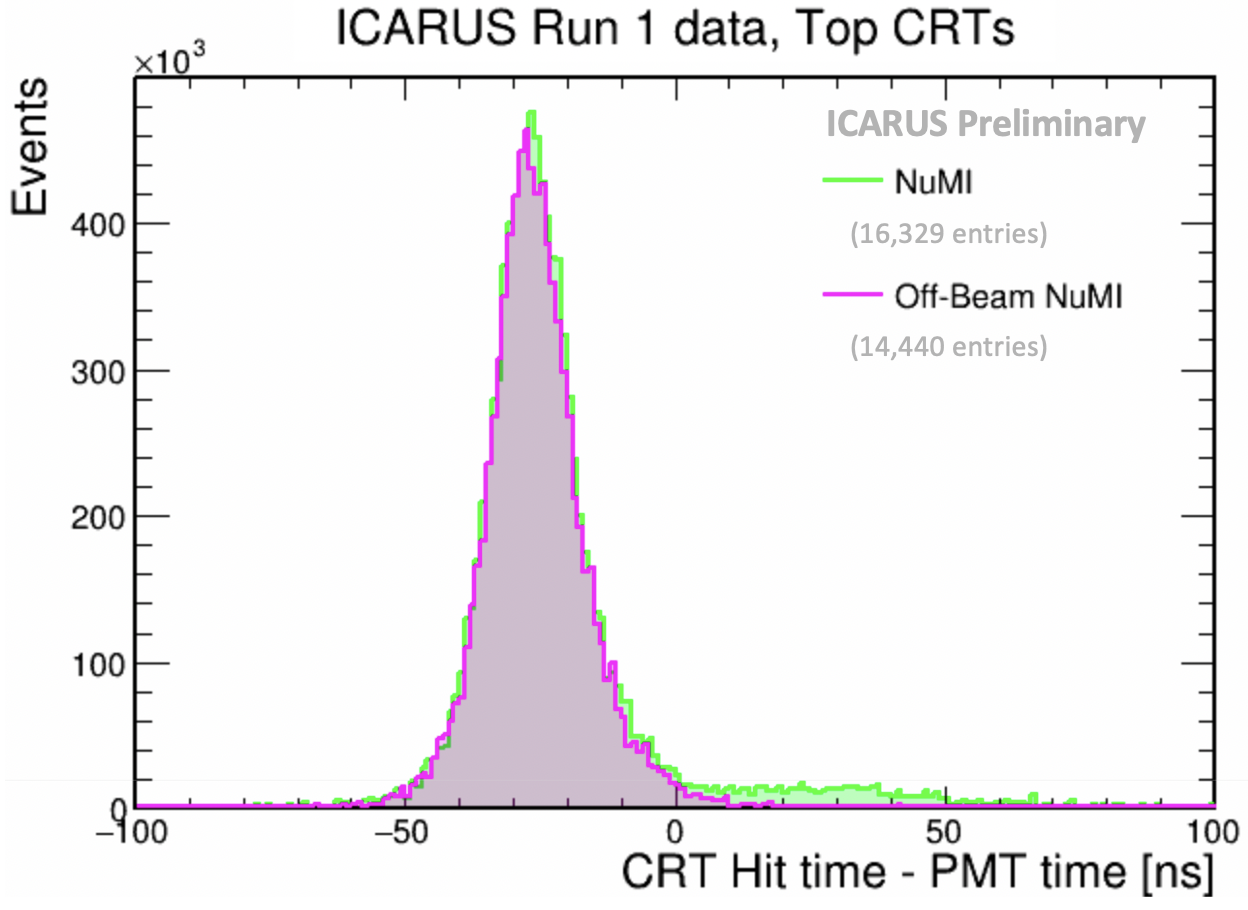}
    \caption{CRT Hit - PMT signal time difference for top CRT hit matches. The distribution is shown for on-beam NuMI data (green) compared with off-beam NuMI data (pink).}
    \label{fig:on-off-beam}
\end{figure}

CRT - PMT matching provides a powerful tool for cosmogenic background rejection based on timing alone, allowing to identify activity that was clearly made from a cosmic. Comparing the off-beam and on-bean data distributions for the top CRT, excess exiting matches are observed with beam-on (see figure \ref{fig:on-off-beam}), demonstrating that CRTs can help disentangle cosmic tracks
from beam activity. 
This method can additionally help examine different topologies of tracks entering and exiting the ICARUS detector. For example, a single PMT flash can be matched to both a top CRT Hit before the flash and a side CRT hit after the flash, and these through-going tracks can be used to study features of the TPC reconstruction. 

\section{Summary}
 ICARUS detector installation and commissioning was completed mid-2022 and collected first physics data from the BNB and NuMI neutrino beams in June 2022. 
 The PMT system is used to identify the interaction time of events, help reconstruct positions of interactions and to generate the trigger for the full detector. CRT system surrounds the detector and tags charged particles entering and exiting the LArTPC. Timing is synchronized across these two subsystems with $\sim$ns level precision, allowing for additional event selection and cosmic rejection.





\acknowledgments
This work was supported by the Office of High Energy Physics within the U.S. Department of Energy Office of Science under Award Number DE-SC0017740.





\end{document}